\documentclass[twocolumn,prb,aps,floatfix,showpacs,superscriptaddress]{revtex4}
\usepackage{graphicx}
\usepackage{dcolumn}
\usepackage{bm}
\usepackage{color}

\begin{document}

\title{Anomalous doping effect in black phosphorene from first-principles calculations}
\author{Weiyang Yu}
\affiliation{International Laboratory for Quantum Functional Materials of Henan, and School of Physics and Engineering, Zhengzhou University, Zhengzhou, 450001, China}
\affiliation{School of Physics and Chemistry, Henan Polytechnic University, Jiaozuo 454000, China}
\author{Zhili Zhu}
\affiliation{International Laboratory for Quantum Functional Materials of Henan, and School of Physics and Engineering, Zhengzhou University, Zhengzhou, 450001, China}
\author{Chun-Yao Niu}
\affiliation{International Laboratory for Quantum Functional Materials of Henan, and School of Physics and Engineering, Zhengzhou University, Zhengzhou, 450001, China}
\author{Chong Li}
\affiliation{International Laboratory for Quantum Functional Materials of Henan, and School of Physics and Engineering, Zhengzhou University, Zhengzhou, 450001, China}
\author{Jun-Hyung Cho}
\email[e-mail address:]{chojh@hanyang.ac.kr}
\affiliation{Department of Physics and Research Institute for Natural Sciences, Hanyang University, 17 Haengdang-Dong, Seongdong-Ku, Seoul 133-791, Korea}
\affiliation{International Laboratory for Quantum Functional Materials of Henan, and School of Physics and Engineering, Zhengzhou University, Zhengzhou, 450001, China}
\author{Yu Jia}
\email[e-mail address:]{jiayu@zzu.edu.cn}
\affiliation{International Laboratory for Quantum Functional Materials of Henan, and School of Physics and Engineering, Zhengzhou University, Zhengzhou, 450001, China}
\date{\today}
\begin{abstract}

\textbf{Abstract}

Using first-principles density functional theory calculations, we investigate the geometries, electronic structures, and thermodynamic stabilities of substitutionally doped phosphorene sheets with group III, IV, V, and VI elements. We find that the electronic properties of phosphorene are drastically modified by the number of valence electrons in dopant atoms. The dopants with even number of valence electrons enable the doped phosphorenes to have a metallic feature, while the dopants with odd number of valence electrons keep a semiconducting feature. This even-odd oscillating behavior is attributed to the peculiar bonding characteristics of phosphorene and the strong hybridization of $sp$ orbitals between dopants and phosphorene. Furthermore, the calculated formation energies of various substitutional dopants in phosphorene show that such doped systems can be thermodynamically stable. These results propose an intriguing route to tune the transport properties of electronic and photoelectronic devices based on phosphorene.


\end{abstract}

\maketitle

\textbf{1.Introduction}

The discovery of graphene has opened many new areas of research related with two-dimensional (2D) atomic-layer systems such as transition metal dichalcogenides (TMDCs), silicene, germanane, and so on.\cite{Geim,Wang,Vogt,Bianco} All of them have attracted much attention as promising materials for future electronics applications.\cite{Butler,Zhang} Most recently, few-layer black phosphorus (BP) was successfully fabricated through exfoliation techniques,\cite{Reich} and especially monolayer BP (termed phosphorene) becomes another stable elemental 2D material. Because of its intriguing electronic properties, phosphorene has drawn much attention of both experimental and theoretical works.\cite{Tran,Fei,Fei2,Tran2,Zhu1,Hu1,Hu2,Manjanath} Interestingly, few-layer BP has been theoretically predicted to have a direct gap or a nearly direct gap ranging from 0.8 to 2 eV depending on the layer thickness.\cite{Tran} It was reported that phosphorene has a high carrier mobility of $\sim$10$^{3}$cm$^{2}$/V$^.$s and an on/off ratio of $\sim$10$^{4}$ at room temperature, thus being considered as a novel channel material in field effect transistors. Recently, Shao \emph{et al} \cite{Shao} predicted that phosphorene would be a potential superconductor material through electron-doping, and Jing \emph{et al} \cite{Jing} investigated the optical properties of phosphorene by molecular doping. These exotic electronic properties of phosphorene can be utilized for the development of future nanoelectronic devices.\cite{Buscema,Li,Ling}

Doping in 2D materials is of fundamental importance to enable a wide range of optoelectronic and electronic devices by tuning their electronic properties. For graphene, it has been well established that carrier concentration can be modulated by charge-transfer doping with adsorbed atoms, molecules, and clusters.\cite{Shin,Thongrattanasiri,Guzmn-Arellano,Huang,Zhang2} Here, the Fermi level can be shifted above or below the Dirac point depending on \emph{n} or \emph{p} doping, respectively.\cite{Tsetseris} Alternatively, substitutional doping in graphene with heteroatoms provides an effective route for simple and stable tuning of doping levels. Although the electronic properties of pure phosphorene have been extensively studied,\cite{Shao,Jing,Buscema,Li,Shin} there are, to our best knowledge, few theoretical and experimental investigations for the effect of various dopants on the electronic properties of phosphorene.

In this work, we perform a systematic study of the substitutional doping of phosphorene with group III, IV, V, and VI elements,respectively. Based on first-principles density functional theory (DFT) calculations, we demonstrate that the electronic properties of phosphorene can be tuned depending on the number of valence electrons in dopant atoms: i.e., group IV and VI elements with even number of valence electrons induce a metallic feature, while group III and V elements with odd number of valence electrons preserve the semiconducting feature of pure phosphorene. The underlying physics of such an even-odd oscillation effect in doped phosphorenes can be explained by the $sp^3$ bonding character of P atoms with a lone pair of valence electrons and their strong hybridizations with the $sp$ orbitals of dopants. Additionally, the thermodynamic stabilities of substitutionally doped phosphorene are also studied by the analysis of formation energy, showing that the doping systems can be realized experimentally.

\textbf{2.Computational method}

\begin{figure}[htb]
\includegraphics[width=7cm]{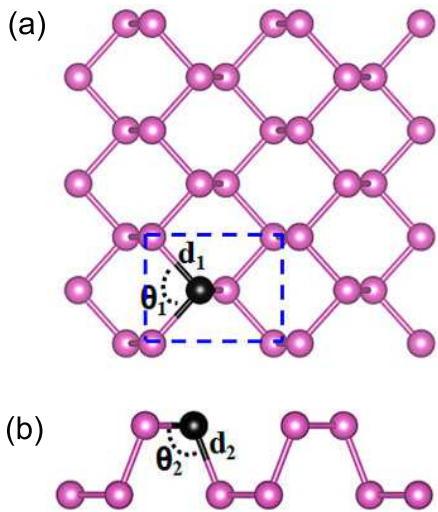}
\caption{(color online). The schematic of phosphorene, (a) top view and (b) side view. The P atom substituted by a dopant atom is drawn with a dark circle. The bond lengths ($d_1$ and $d_2$) and bond angles (${\theta}_1$ and ${\theta}_2$) are represented in Table I.}
\end{figure}

Our DFT calculations within the general gradient approximations (GGA) have been performed using Vienna \emph{ab initio} simulation package (VASP) code.\cite{Kresse} We used the Perdew-Burke-Ernzerhof (PBE)\cite{Perdew} exchange-correlation functional for the GGA. The projector augmented wave (PAW) method\cite{Kresse2} was employed to describe the electron-ion interaction. In the structural optimization, all the atoms in the modeling systems were allowed to relax until all the residual force components were less than 0.01 eV/{\AA}. For the calculations of the density of state (DOS), tetrahedron method is used with a quick projection scheme. For the calculations of the band structures, we use Gaussian smearing in combination with a small width of 0.05 eV, and the path of integration in first Brillouin zone is along Y(0.0, 0.5, 0.0)$\rightarrow$ $\Gamma$(0.0, 0.0, 0.0)$\rightarrow$X(0.5, 0.0, 0.0)$\rightarrow$M(0.5, 0.5, 0.0). A kinetic energy cutoff of 500 eV was used in all calculations.

The present substitutionally doped phosphorene systems were modeled in a periodic slab geometry, where a series of 2$\times$2 and 4$\times$4 supercells of the phosphorene sheet was used with a vacuum spacing of $\sim$15 \AA\ between adjacent phosphorene sheets, respectively. Here, one P atom in 2$\times$2 or 4$\times$4 supercells was substituted by a dopant atom, as shown in figure 1(a) and 1(b), leading to a doping concentration of 6.25\% and 1.56\%, reapetively. The $k$-space integration was done with 8$\times$10$\times$1 and 4$\times$4$\times$1 ${\bf k}$ points in the 2$\times$2 and 4$\times$4 surface Brillouin zone, respectively, generated by the Monkhorst-Pack scheme.\cite{Monkhorst} To test our results, we have used relatively larger larger supercells up to 5$\times$5 and 6$\times$6 for all the dopants with smaller concentration of 1.00\% and 0.69\%, respectively, and we obtained the same trend. Therefore, we here present our results of 2$\times$2 or 4$\times$4 supercells. Finally, for the 4$\times$4$\times$2 supercell for bulk doping calculation and 4$\times$8 supercell for two elements codoping, the ${\bf k}$ points of 3$\times$3$\times$3 and 2$\times$2$\times$1 were applied, respectively. The distance of two dopants in the codoping systems is set to 4.33{\AA} in order to model the random distribution of dopant.

\begin{table}
\caption{Calculated bond lengths ($d_1$ and $d_2$) and bond angles (${\theta}_1$ and ${\theta}_2$) for 2$\times$2 supercell of B, C, N, O, Al, Si, S, Ga, Ge, As and Se doped phosphorenes, respectively, together with those of pure posphorene. For comparison, previous theoretical results for pure phosphorene are also given.}
\begin{tabular}{p{2cm}p{1.4cm}p{1.4cm}p{1.4cm}p{1.3cm}}
\hline
\hline
Dopant atom & $d_{1}$ (\AA)& $d_{2}$ (\AA) & $\theta_{1}$ ($^\circ$)& $\theta_{2}$ ($^\circ$)  \\
\hline
pure           & 2.24      & 2.26        &  96.6         &  101.6   \\
pure~\cite{Du} & 2.25      & 2.26        &  96.9         &  102.3  \\
B              & 1.95      & 1.88        &  108.7        &  122.3  \\
C              & 1.81      & 1.80        &  107.2        &  120.6   \\
N              & 1.80      & 1.80        &  104.9        &  119.0   \\
O              & 2.11      & 1.75        &  110.4        &  117.1  \\
Al             & 2.36      & 2.35        &  97.3         &  118.2  \\
Si             & 2.27      & 2.30        &  96.0         &  106.8  \\
S              & 2.18      & 2.97        &  99.1         &  103.3 \\
Ga             & 2.36      & 2.35        &  100.5        &  119.8  \\
Ge             & 2.26      & 2.26        &  96.2         &  103.9  \\
As             & 2.35      & 2.39        &  93.9         &  104.3    \\
Se             & 2.32      & 3.02        &  99.1         &  94.6    \\
\hline
\hline
\end{tabular}
\end{table}

\begin{figure*}[htb]
\includegraphics[width=12cm]{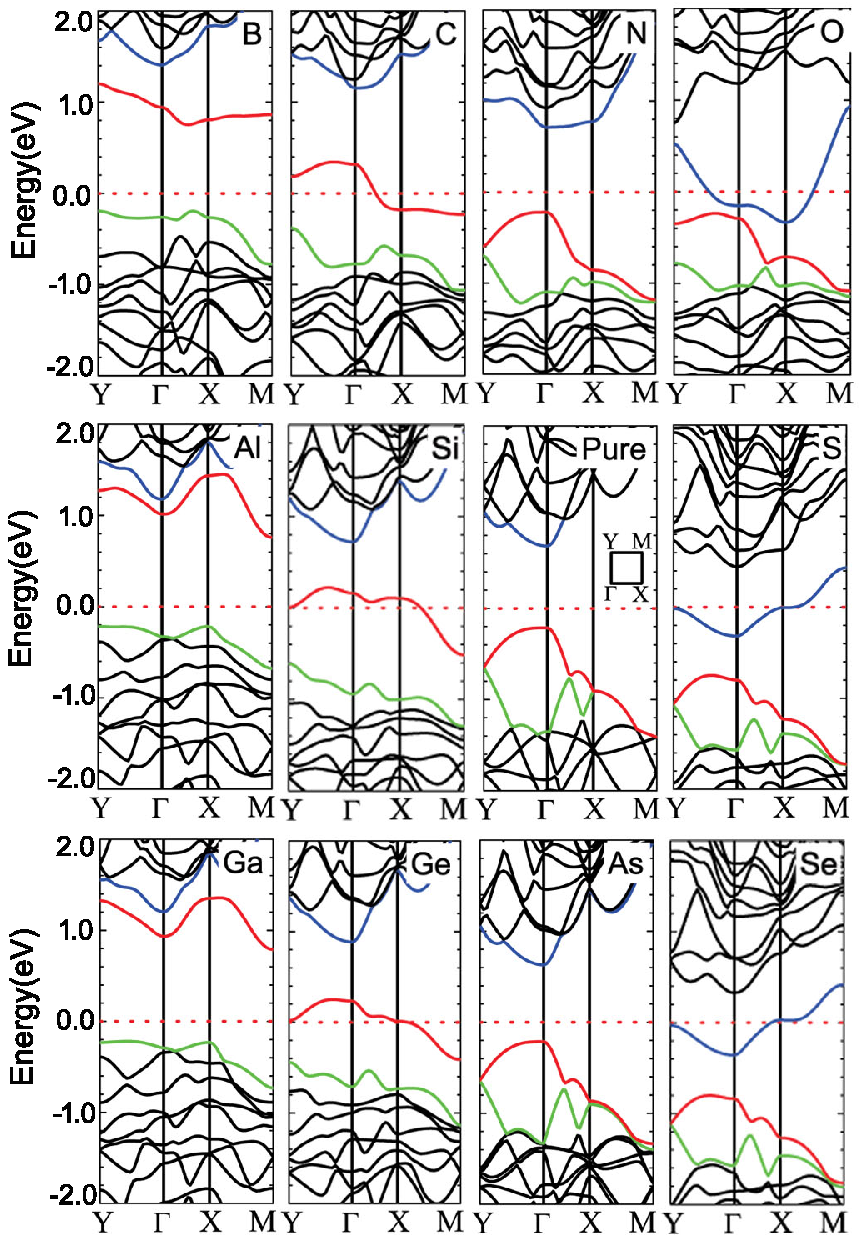}
\caption{(color online) Calculated band structures for 2$\times$2 supercell with various dopants in phosphorenes from group III, IV, V, and VI, respectively, together with that of pure phosphorene. The band dispersions are plotted along the symmetry lines shown in the surface Brillouin zone (see the inset). The energy zero represents the Fermi level.}
\end{figure*}

\begin{figure*}[htb]
\includegraphics[width=12cm]{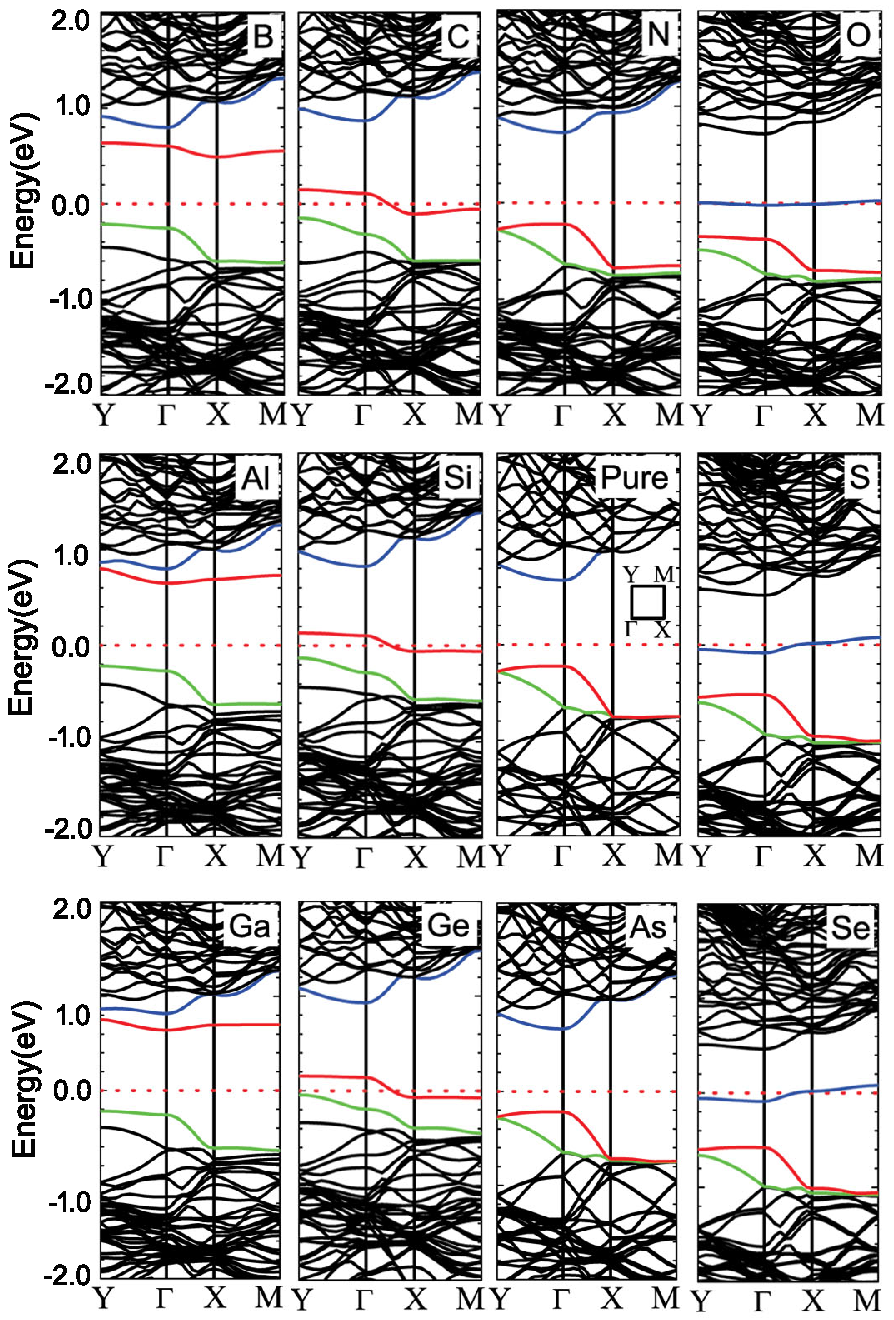}
\caption{(color online) Calculated band structures for 4$\times$4 supercell with various dopants in phosphorenes from group III, IV, V, and VI, respectively, together with that of pure phosphorene. The band dispersions are plotted along the symmetry lines shown in the surface Brillouin zone (see the inset). The energy zero represents the Fermi level.}
\end{figure*}

\begin{figure}[htb]
\includegraphics[width=8.3cm]{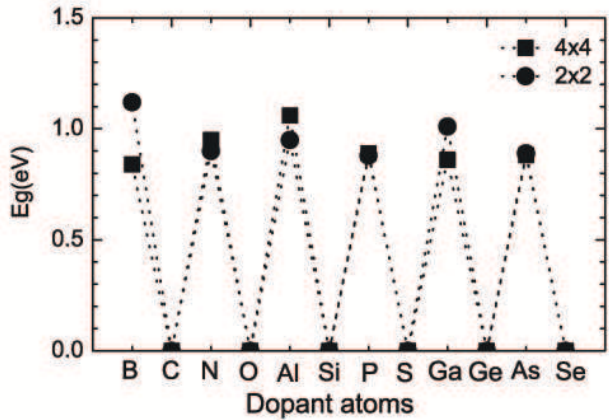}
\caption{Calculated band gaps of substitutionally doped phosphorenes with various dopants form group III, IV, V, and VI, respectively.}
\end{figure}

\begin{figure}[htb]
\includegraphics[width=8.5cm]{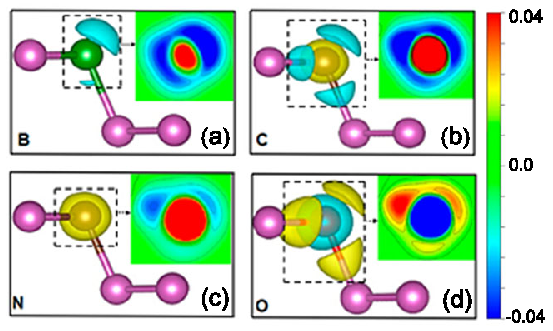}
\caption{ (color online). Schematic diagrams of difference charge densities $\Delta$$\rho$ = $\rho$$_{doped}$ - $\rho$$_{pure}$ of (a) B, (b) C, (c) N and (d) O doped phoshporenes, respectively. Iso surfaces correspond to 0.04 e/{\AA}$^{3}$.}
\end{figure}

\textbf{3.Results and discussions}

To obtain the stable geometry structures of the doped systems, we optimize the atomic structure of each substitutionally doped phosphorene. The calculated geometric parameters were listed in the Table I, the the bond lengths, denoted as $d_1$ and $d_2$ , and bond angles, ${\theta}_1$ and ${\theta}_2$ in figure 1(a) and 1(b), respectively. The dopants here we study are B, C, N, O, Al, Si, S, Ga, Ge, As and Se from group III to group VI.  The geometric parameters of pure phosphorene were also listed for comparison. From our DFT calculations, these geometric parameters of pure phosphorene agree well with those obtained by a previous calculation\cite{Du}(see Table I). In the doped phosphorence, we find that the $d_1$ and $d_2$ (${\theta}_1$ and ${\theta}_2$) values for B, C, N, and O dopants are smaller than the corresponding ones of pure phosphorene, because of the much smaller atomic radii of B, C, N and O atoms compared with P atom. While the geometric parameters in Al, Si and S doped phosphorenes are close to that of pure phosphorene because of the same period with P atom in Element Periodic Table, and the structure parameters in Ga, Ge, As and Se doped phosphorenes are larger than that of pure phosphorene on account of the larger atomic radii.

Now, we begin to investigate the electronic structures of the doped phosphorence. We first chose a 2$\times$2 supercell in which one P atom is replace by dopant to simulate the larger doping concentration, namely 6.25\%. The calculated band structures of B, C, N, O, Al, Si, S, Ga, Ge, As and Se doped phosphorenes, together with that of pure phosphorene, is plotted in Figure 2. From the band structure of pure phosphorence, we find that the valence band is occupied while the conduction band unoccupied with a band gap about 1 eV in our calculation,which is in agreement with other previous calculations\cite{Du}. For the doped systems, we can see that the Fermi levels in C, Si, Ge and B, Al, Ga doped phosphorenes shift downward toward the valence band while that in O, S, Se doped phosphorene shift upward toward the conduction band, as shown in figure 2. More interestingly, the electronic properties of doped phosphorene systems are determined dramatically by the number of valence electrons in dopant atoms:
i.e., C, Si, Ge and O, S, Se atoms with even number of valence electrons give rise to metallic features while doping atoms with odd number of valence electrons (such as B, Al, Ga and N, As) remain semiconducting.

To investigate whether the above metallic-semiconducting oscillation with number of the valence electrons is general, we calculate the band structures of various dopants using a relative larger 4$\times$4 supercell with a smaller concentration of 1.56\%. The results are presented in figure 3. From figure 3, the even-odd oscillation behaviors also can be found as the same as in figure 2. Beside this, we noticed that such interesting properties is still preserved by reducing the doping concentration up to 1\% or even less than this, i.e. some 5$\times$5 and 6$\times$6 supercells were used in the calculations. Of course, if the doping concentrations become very small, the even-odd oscillations will be disappear undoubtedly. Therefore, in the following we focus our discussions on a relative high doping concentration.

We summarize the general trend of the above-mentioned even-odd oscillations behaviors for 2$\times$2 and 4$\times$4 supercells in figure 4. The oscillating schematic diagram in figure 4 demonstrates the calculated band gaps ($E_g$) of various dopants in phosphorene. From figure 4 we can see that for 4$\times$4 supercell, the band gap of B, Al, Ga (N, As) doped phosphorene are 0.84, 1.12, 0.89 (0.95, 0.88) eV, respectively, while the band gap of pure phosphorene is 0.90 eV, which is consistent with the results of Liang \emph{et al}\cite{Liang}. For 2$\times$2 supercell the band gap of B, Al, Ga (N, As) doped phosphorene are 1.12, 0.93, 1.05 (0.90, 0.89) eV, which is almost the same as that of 4$\times$4 supercells.
It is seen that the band gap oscillates depending on even and odd numbers of valance electrons in dopant atoms. Here, group IV (C, Si and Ge) or group VI (O, S and Se) doped phosphorenes that have odd number of electrons per unit cell exhibit a metallic feature with a half-filled band at the Fermi level. On the other hand, group III (B, Al and Ga) or group V (N and As) doped phosphorenes containing even number of electrons per unit cell have fully occupied valence bands, showing a semiconducting feature.
Generally speaking, the decrease(increase) of gap opening in doped phosphorenes is most likely to be associated with the strong hybridization of $p$ orbitals between dopants and phosphorene, as discussed below.

\begin{figure}[htb]
\includegraphics[width=8.5cm]{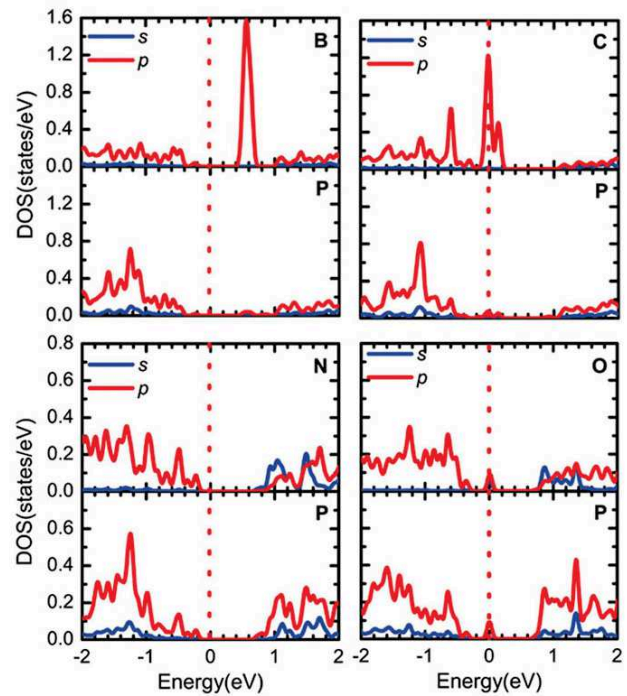}
\caption{ (color online). Partial DOS for B, C, N and O doped phosphorenes. The partial DOS projected onto the 2$s$ and 2$p$ orbitals of the dopant and its bonding P atom are displayed. The energy zero represents the Fermi level.}
\end{figure}

\begin{figure}[htb]
\includegraphics[width=8.5cm]{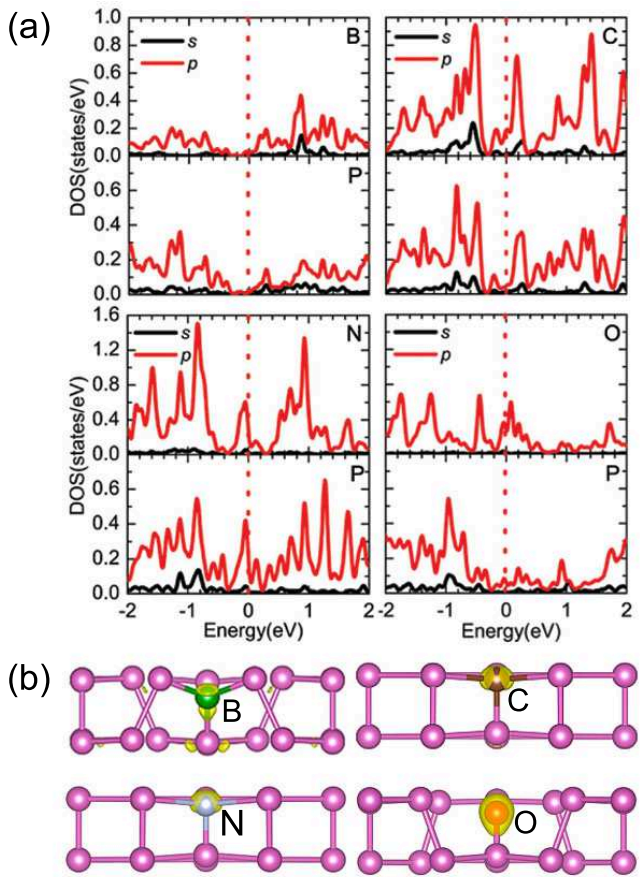}
\caption{(color online). (a) Density of states of substitutional B, C, N, and O doped bulk black phosphorus. Energies refer to the Fermi energy. (b) Side view of band decomposed charge density of bulk black phosphorus with energy from -1 to 1 eV around Fermi level. Iso surfaces correspond to 0.02 e/{\AA}$^{3}$.}
\end{figure}

The presence of metallic and semiconducting properties in doped phosphorenes can be associated with the bonding nature of phosphorene where
each P atom (having valence electron configuration 3$s^2$3$p^3$) forms $sp^3$ bonding with a lone pair of valence electrons.\cite{Rudenko}
In doped phosphorenes, the non-bonding lone-pair electronic states of pure phosphorene are easily tuned by the valence-electron states of dopant atoms, leading to the semiconducting or metallic electronic states near the Fermi level, as shown in figures 2 and 3. Furthermore, we plot the difference charge densities between doped system and pure phosphorene in figure 5 for B, C, N and O-doped systems, respectively. From figure 5 we can see that for N-doped phosphorene, three electrons of N involved in the bonding between N and neighboring P atoms, leaving one lone pair electrons on each atom, which make the N-doped system keep the same semiconducting property as pure phosphorene. While for B-doped phosphorene, the lone pairs on B atom disappear because of the only three valance electrons bonding to neighboring P atoms, keeping also semiconducting property. As for C-doped phosphorene, there is only one electron not involved in, leading to a half-filled delocalized energy band. While for O-doped phosphorene, three non-bonding electrons also give rise to half-filled state, presenting metallic properties. Our findings of the metal-semiconductor oscillatory behaviors in doped phosphorene systems contrast with the conventional doping effect in group IV semiconductors, where $\emph{n}$- and $\emph{p}$-type dopants create some localized doping-induced states in band gap.

\begin{figure}[htb]
\includegraphics[width=8.5cm]{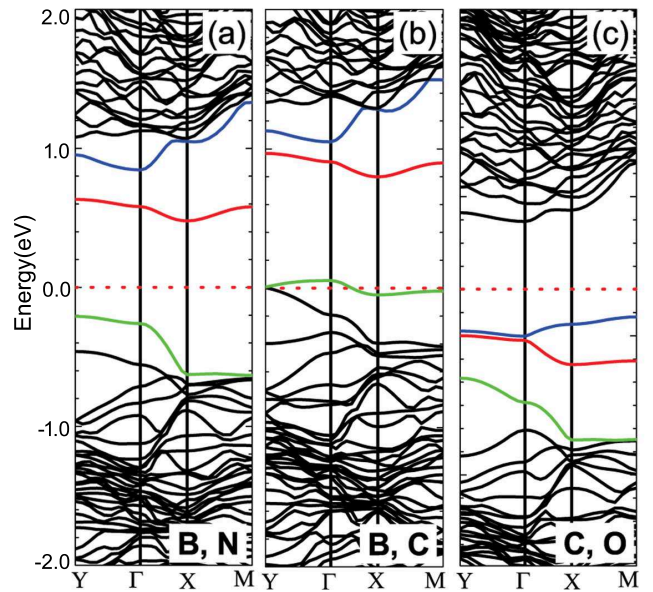}
\caption{ (color online). Calculated band structures of (a) B and N codoped phosphorene, (b) B and C codoped phosphorene, (c) C and O codoped phosphorene, respectively. The energy zero represents the Fermi level.}
\end{figure}

To further shed light on the underlying bonding mechanism of the dopants and the phosphorus atoms, we show in figure 6 the total and partial density of states (PDOS) of B, C, N and O-doped phosphorene, respectively. As shown in figure 6, the calculated partial density of states (DOS) projected onto the 2$p$ orbitals of each dopant atom shows similar pattern and peak positions as compared with that for P atom, indicating a strong hybridization of $sp$ orbitals between dopants and phosphorene. This strong hybridization leads to a broadening of the partial DOS for dopants. Especially, for C (O) doped phosphorene, the $2p$ partial DOS peaks for the C (O) dopant and P atom locate at the Fermi level, thereby giving rise to a metallic property. We have double-checked the results with spin-polarized calculation especially for the metallic behavior of dopants with even number of valence electrons. The spin-polarized calculations show that there is no spin split at Fermi level. So, the results of metal properties for dopants with even number of valence electrons are proper.

On the other hand, compared with the electronic properties of the monolayer doped phosphorene, the odd-even oscillations behavior in doped bulk black phosphorus does not exist anymore. The reason is that, owing to the van der waals interaction between the interlayers in bulk black phosphorus, there is no lone pair electrons on the atoms of neither the pure nor doped black phosphorus. This point can simply be illustrated from the DOS of the doped bulk black phosphorus. In figure 7 (a), we plot the DOS of B-, C-, N-, and O-doped bulk black phosphorus with the concentration of 0.39\% for a 4$\times$4$\times$2 supercell simulations. From figure 7 (a) we can see that the peaks of DOS exit at the Fermi level, both in the B(N)-doped systems and in the C(O)-doped systems. This suggests there may be impurity states in the doping systems. Indeed, the impurity states originate from the dopant atoms, as shown in figure 7 (b). In all, it is different from the two dimensional doping systems because of the low dimensional size effect, as mentioned above.

\begin{figure}[htp]
\includegraphics[width=8.3cm]{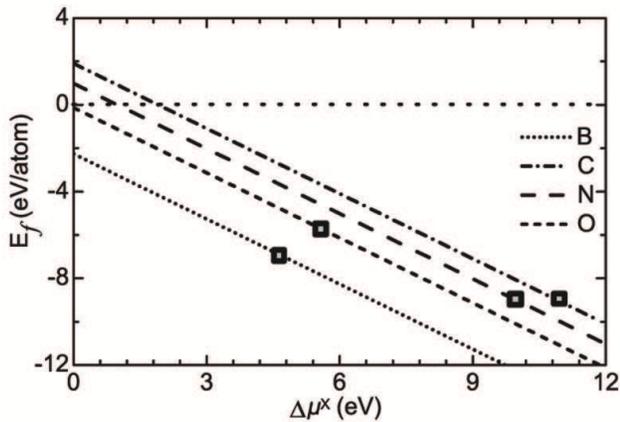}
\caption{Calculated formation energies of substitutional B, C, N and O dopings in phosphorene as a function of the chemical potential difference of each dopant atom. The reference zero of ${\Delta}{\mu}^{\rm X}$ represents the chemical potential obtained from orthorhombic B bulk, graphite, N$_{2}$ and O$_{2}$, respectively. The formation energies estimated from the atomic chemical potential of each dopant are marked with squares.}
\end{figure}

Our above new findings provide a new approach to tune the electronic properties by dopants with two types of different elements and different number of valence electrons. For example, we can tune the 2D phosphorene by doping with two different dopant atoms, namely co-doping\cite{Zhu2}. We perform a 4$\times$8 supercell calculation in which two P atoms were replaced by two different elements, such as B and N, B and C, C and O, respectively. The co-doped phosphorene systems exhibit nontrivial behavior, the calculation results are shown in figure 8. The band structure of B and N co-doped phosphorene is shown in Figure 8(a). It is naturally that B and N co-doped system displays a semiconductor property. As for B and C co-doped phosphorene, as shown in figure 8(b), the band structure exhibits metal property because of the odd and even valence electron. While for C and O co-doped phosphorene, as shown in figure 8(c), the band structure also exhibits a semiconductor property. It is not difficulty to understand that the valence band of C and O co-doped phosphorene is fully occupied, leading to a semiconductor property.

Finally, it is reasonable to question that whether such doping system can be realized experimentally. To meet this query, we examine the thermodynamic stability of doped phosphorene by calculating the formation energy of substitutional doping which is defined as E$_{f}$ = (E$_{tot}^{\rm P+X}$ + $\mu^{\rm P}$) $-$ (E$_{tot}^{\rm P}$ + $\mu^{\rm X}$). Here, E$_{tot}^{\rm P+X}$ and E$_{tot}^{\rm P}$ denote the total energies of X-doped phosphorene and pure phosphorene, respectively; $\mu^{\rm P}$ is the chemical potential of phosphorus (taken from pure phosphorene) while $\mu^{\rm X}$ is the chemical potential of X dopant atom. The calculated formation energies of B, C, N, and O doped phosphorenes are plotted as a function of ${\Delta}\mu^{\rm X}$ in figure 9. We find that, if we take $\mu^{\rm X}$ from the energies of orthorhombic B bulk, graphite, O$_2$, and N$_2$, the formation energies of C and N (B and O) doped phosphorenes are positive (negative), indicating an energy cost (gain) for the substitutional doping. However, all the formation energies become negative (marked with squares in figure 9) by assuming $\mu^{\rm X}$ at their respective upper limits, i.e., at their atomic energies. Thus, we can say that the substitutional B and O dopings can be easily realized in experiments, but the substitutional C and N ones need to be particularly cautious in the synthesis processes such as atomic deposition rate, temperature, and other experimental conditions.

\textbf{4. Conclusions}

In conclusion, the electronic properties of substitutionally doped phosphorenes with group III, IV, V, and VI elements have been systematically investigated by first-principles DFT calculations. We found that group IV and VI dopants with even number of valence electrons produce the metallic property whereas group III and V dopants with odd number of valence electrons preserve the semiconducting property. This even-odd behavior of the electronic properties of doped phosphorenes is revealed to be due to the strong hybridization of $\emph{sp}$ orbitals between dopants and phosphorene. The estimated formation energies of the substitutional B, C, N, and O doped phosphorenes provide an information for their thermodynamic stabilities, which can be realized in experiments. Our new findings of substitutionally doped phosphorenes are drastically different from conventional $\emph{n}$- or $\emph{p}$-type doping effect in group IV semiconductors. The novel metal-semiconductor oscillations predicted here not only provide an intriguing route to tune the transport properties of electronic devices based on phosphorene materials, but also stimulate experimentalists to develop new phosphorene-based nanoelectronic devices.

\textbf{Acknowledgments}

We thank Prof. Zhenyu Zhang for helpful discussions. This work was supported by the National Basic Research Program of China (Grant No. 2012CB921300), National Natural Science Foundation of China (Grant Nos. 11274280 and 11304288), and National Research Foundation of Korea (Grant No. 2014M2B2A9032247).

\end{document}